\documentclass[12pt]{article}
\usepackage{amsmath}
\usepackage{amssymb}
\usepackage{epsfig}

 \advance \textwidth by 0.1cm
 \advance \textheight by 0.06cm
\def\##1{{\underline #1}}
\def\~#1{{\underline {\mathcal #1}}}
\def\+#1{{{\mathcal #1}}}
\def\=#1{\underline{\underline #1}}

\def\.{\mbox{ \tiny{$^\bullet$} }}

\begin{document}

\begin{center}{\large\bf ON WIDENING THE ANGULAR EXISTENCE DOMAIN FOR DYAKONOV SURFACE WAVES USING THE POCKELS EFFECT }\end{center} \vskip 1 cm

\noindent SUDARSHAN R. NELATURY\footnote{Corresponding author}\\
Department of Electrical, Computer and Software Engineering, \\
Pennsylvania State University, The Behrend College,\\
5101 Jordan Road, Erie, PA 16563-1701, USA.\\

\noindent JOHN A. POLO JR.\\
Department of Physics and Technology,\\
Edinboro University of Pennyslvania,\\
235 Scotland Rd., Edinboro, PA  16444, USA.\\

\noindent AKHLESH LAKHTAKIA\\
CATMAS~---~Computational and Theoretical Materials Sciences Group, \\
Department of Engineering Science and
Mechanics, \\
Pennsylvania State University, University Park, PA  16802, USA.

\begin{abstract}
{The propagation of Dyakonov surface waves (DSWs) at the planar interface
between an isotropic material and a linear electro-optic birefringent material
can be dynamically controlled using the Pockels effect. The
range of directions for DSW propagation has been previously found
to be rather narrow. By careful choice of various parameters, this
range of directions can be increased by more
than an order of magnitude.}


\end{abstract}

\section{Introduction}
Electromagnetic surface waves   at the planar interface of two
different materials travel along a direction
% lying in \footnote{\textbf{Akhlesh and Sudarshan, would "parallel to" be better than "lying in"?}}
%the interface\footnote{\textbf{Akhlesh and Sudarshan, would "interfacial" be better?}} plane
parallel to the interfacial plane with decaying amplitudes
perpendicular to the direction of propagation in both materials.
Research on surface-wave propagation  can be traced back to about a
century ago to Zenneck~\cite{Zenneck} and continues to be of
technoscientific importance~\cite{PLeditorial}.

In recent years, Dyakonov surface waves (DSW) have attracted much attention~\cite{Takayama08}.
 These waves can exist at the planar interface of an isotropic material and
 %an\footnote{\textbf{Akhlesh and Sudarshan, shouldn't this be "a" rather than "an"?}} uniaxial material,
a uniaxial material, both dielectric, as was shown first by
D'yakonov \cite{Dyakonov} and then enlarged upon by Averkiev and
Dyakonov \cite{Averkiev}. The isotropic material is not metallic.
Later, several researchers considered DSW propagation on
increasingly complex bimaterial interfaces such as
biaxial-isotropic, uniaxial-uniaxial, and biaxial-biaxial
materials---see, e.g., Refs.~6--10.

DSW propagation is assured when a  dispersion equation is satisfied.
This equation arises from having to satisfy the frequency-domain
Maxwell postulates in both materials as well as the boundary
conditions at the interface. Therefore, DSW propagation depends
highly on the crystallographic symmetries of the two materials, at
least one of which must be birefringent. Calculations for many pairs
of materials have shown that conditions for DSW propagation are
favored only for a narrow and restricted angular range of
propagation directions in the interface
\cite{Takayama08,Polo_06_2,Walker,Crasovan06,Artigas05,Crasovan05}.
The width of this  angular existence domain (AED) is denoted here by
$\Delta\psi$. With the exception of some interfaces involving
hypothetical metamaterials~\cite{Artigas05}, $\Delta\psi$ for
interfaces of non-magnetic materials are small~---~in most cases,
only a fraction of a degree\cite{Takayama08}.  This makes
experimental observation of DSW propagation  very difficult.  One of
the largest values of $\Delta\psi$ has been calculated to be
$\approx 4^\circ$~\cite{Takayama08}; however, in that case one of
the materials is calomel, a chalky and dusty material that is
impractical to use.

The potential real-life applications envisaged for the use of DSW
are numerous~\cite{Takayama08}, including devices for integrated
optics, chemical and biological surface sensing, etc. To be able to
experimentally launch a DSW, to experimentally detect a DSW, and to
possibly control the DSW characteristics, a sufficiently wide AED
appears desirable. That desire motivated this communication.

 In a predecessor paper~\cite{NPL07_1}, we looked at the control of DSW propagation at the interface of an electro-optic material, potassium niobate, and an isotropic material through application of an external dc electric field.  Now, we examine
 optimizing the width $\Delta\psi$ of the AED by an exploration of the effects of:  the crystal orientation,
 the direction of applied dc electric field, and
 the strength of the applied dc electric field. Considering the planar
interface of an isotropic and a birefringent material,
both dielectric, we
show in Sec.~\ref{eo} that $\Delta\psi$ can be increased by a
large factor---if the birefringent material
is susceptible to the Pockels effect---by suitable choices of parameters
such as the magnitude and orientation of an applied dc electric field,
the crystallographic orientation of the birefringent material
in relation to the interface, and
the refractive index of the isotropic material.

\section{Preliminaries and Numerical Results}\label{eo}

We considered the following problem \cite{NPL07_1}.   The half-space $z< 0$ is filled with a homogeneous,
isotropic, dielectric material with an optical refractive index
denoted by $n_s$. The half-space $z> 0$ is filled with a
homogeneous, linear, electro-optic material, whose optical relative
permittivity matrix  is stated as
\begin{equation}
\bar{\epsilon}_{\,rel} = \bar{S}_{z}\left(\psi\right)\cdot
\bar{R}_{y}(\chi) \cdot\bar{\epsilon}_{PE}
\cdot\bar{R}_{y}(\chi)\cdot \bar{S}_{z}\left(-\psi\right)\,.
\label{AAepsr}
\end{equation}
Incorporating the Pockels effect due to an arbitrarily oriented  but
uniform dc electric field ${\#E}^{dc} $, the matrix
$\bar{\epsilon}_{PE}$  is given by
\begin{equation}
\displaystyle{\bar{\epsilon}_{PE}\approx\left(
\begin{array}{ccc}
\epsilon _{1}^{(0)}(1-\epsilon _{1}^{(0)}s_1 ) & -\epsilon _{1}^{(0)}\epsilon
_{2}^{(0)}s_6 &
-\epsilon _{1}^{(0)}\epsilon _{3}^{(0)}s_5 \\[5pt]
-\epsilon _{1}^{(0)}\epsilon _{2}^{(0)} s_6
& \epsilon _{2}^{(0)}(1-\epsilon _{2}^{(0)}s_2 ) &
-\epsilon _{2}^{(0)}\epsilon _{3}^{(0)} s_4 \\[5pt]
-\epsilon _{1}^{(0)}\epsilon _{3}^{(0)} s_5
& -\epsilon _{2}^{(0)}\epsilon _{3}^{(0)} s_4 & \epsilon _{3}^{(0)}(1-\epsilon
_{3}^{(0)} s_3 )
\end{array}
\right) }\,,  \label{PocEps}
\end{equation}
correct to the first order in $\vert\#E^{dc}\vert$, where
\begin{equation}
s_j=\sum_{K=1}^3
r_{jK}E_{K}^{dc}\,, \qquad j\in[1,6]\,,
\end{equation}
\begin{equation}
\left(
\begin{array}{l}
E_{1}^{dc} \\[5pt]
E_{2}^{dc} \\[5pt]
E_{3}^{dc}
\end{array}
\right) = \bar{R}_{y}(\chi)\cdot\bar{S}_{z}\left(-\psi\right)\cdot
\left(
\begin{array}{l}
E_{x}^{dc} \\[5pt]
E_{y}^{dc} \\[5pt]
E_{z}^{dc}
\end{array}
\right) \,,
\end{equation}
$\epsilon _{1,2,3}^{(0)}$ are the principal relative permittivity
scalars in the optical regime, whereas $r_{JK}$ (with $1\leq J\leq
6$ and $1\leq K\leq 3 $) are the electro-optic coefficients. The electro-optic material
can be  isotropic, uniaxial, or biaxial, depending on the
relative values of $\epsilon_1^{(0)}$, $\epsilon_2^{(0)}$, and $%
\epsilon_3^{(0)}$. Furthermore, this material may belong to one of
20 crystallographic classes of point group symmetry, in accordance
with the relative values of the  coefficients $r_{JK}$.

The rotation matrix
\begin{equation}
\bar{S}_z(\psi)=\left(
\begin{array}{ccc}
\cos \psi & -\,\sin\psi & 0 \\
\sin\psi & \cos \psi & 0 \\
0 & 0 & 1
\end{array}
\right)
\end{equation}
in Eq.~(\ref{AAepsr}) denotes a rotation about the $z$ axis by
an angle $\psi \in\left[0,2\pi\right)$. The matrix
\begin{equation}
\bar{R}_{y}(\chi )=\left(
\begin{array}{ccc}
-\sin \chi & 0 & \cos \chi \\
0 & -1 & 0 \\
\cos \chi & 0 & \sin \chi
\end{array}
\right)
\end{equation}
involves the angle $\chi \in\left[0,\pi/2\right]$ with respect to
the $x$ axis in the $xz$ plane, and combines a rotation as well as
an inversion.  The angles $\psi$ and $\chi$ delineate the
orientation of the electro-optic material in the laboratory coordinate system,
the full transformation from laboratory coordinates ($x,y,z$) to
those used conventionally for electro-optic materials ($1,2,3$) being
illustrated by Nelatury \emph{et al.}~\cite{NPL07_1}.

The wave vectors on both sides of the interface were taken to lie
wholly in the $xz$ plane.
The field representations based on the Maxwell curl postulates in
the two half-spaces, the boundary conditions at the interface $z=0$,
and the dispersion equation for DSW propagation are available
elsewhere~\cite{NPL07_1}. For a simple case like that of the
isotropic-uniaxial interface, obtaining a closed-form expression for
$\Delta\psi$ is relatively easy. However, for complex cases such as the
one we are treating here, that goal is perhaps impossible.
Recourse, then, has to be made to numerical
methods in order to mount an extensive search for the proper choice
of parameters that allow DSW propagation.

A search was carried to maximize $\Delta\psi$ in relation to the
magnitude and the direction of the dc electric field, the refractive
index of the substrate, and the tilt angle $\chi$. Potassium niobate
was chosen as the electro-optic material, because   the magnitudes of its
electro-optic coefficients are very large.

\begin{figure}
    \begin{center}
    \includegraphics[width=14cm]{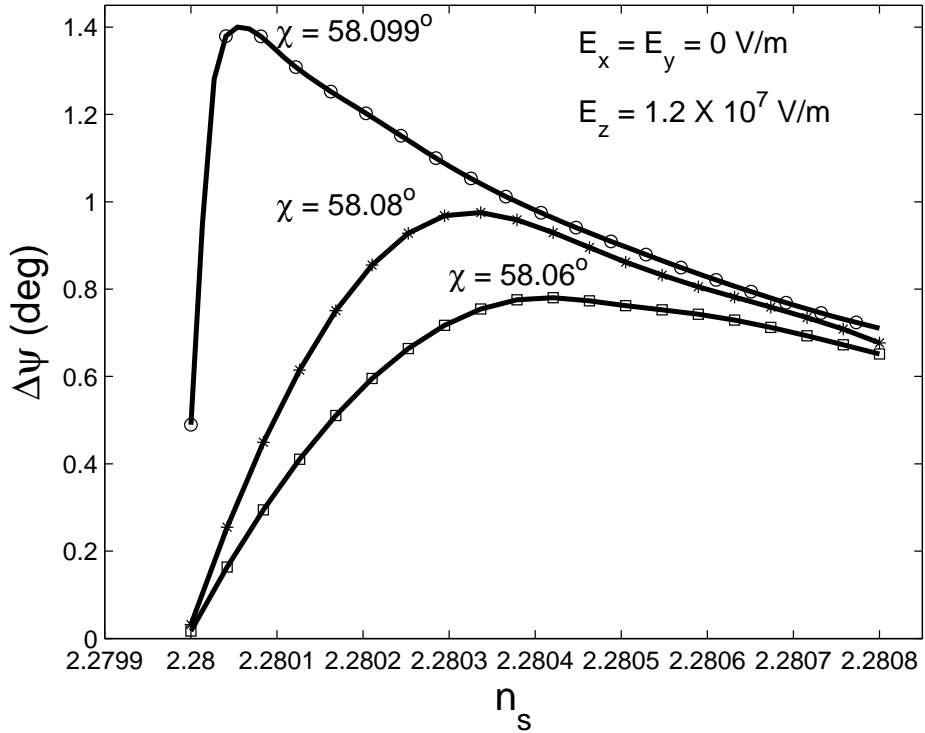}
    \end{center}
    \caption{   Variation of $\Delta\psi$ versus $n_s$ for different $\chi$ values }
    \label{Fig:Delta_Psi_curves}
 \end{figure}

Table 1 shows $\chi$ and $n_s$
values in a neighborhood wherein $\Delta\psi$ is a maximum.
Figure~\ref{Fig:Delta_Psi_curves} shows the variation of
$\Delta\psi$ with $n_s$ for  $\chi = 58.06^\circ, 58.08^\circ $ and
$58.099^\circ$. The dc electric field is
${\#E}^{dc}  =\hat{z}\, 1.2\times10^7 $~V~m$^{-1}$.
We found that, around $\chi= 58^\circ$, even a small change in
$\chi$ produces a large change in $\Delta\psi$. The last entry in
Table 1 shows $\Delta\psi\approx 1.4^\circ$ which is 20 times larger
than the result previously reported by us \cite{NPL07_1}.

The search for a wider AED was also attempted choosing ${\#E}^{dc}$
parallel to $-\hat{z}$, $\hat{x} \cos\psi +\hat{z}\sin\psi$ and
$-\hat{x} \sin\psi +\hat{z}\cos\psi$ for the same choices of $n_s$
and $\chi$ values. But we could get only $\Delta\psi$ of order
$0.02^\circ$. However, we wish to continue the search hoping for
wider AED.

\section{Concluding Remarks}

In summary, Dyakonov surface waves propagate along interfaces
involving birefringent materials over a narrow angular range
$\Delta\psi$ of the orientation angle $\psi$. Having a larger
$\Delta\psi$ is helpful for easier generation and detection of DSWs.
Several attempts have been made to widen this angular existence
domain. By considering specific directions of an applied dc electric
field and specific values of $\chi$, we have shown that  exploitation of the linear
electro-optic (Pockels) effect has the potential to widen the AED.
The search for higher values of $\Delta\psi$ for an arbitrarily
oriented dc electric field is still open. Also, one might look into ranges of $\chi$
values for specific values of $\psi$ and various magnitudes and
directions of the applied dc electric field.

\begin{table}
\begin{center}
\begin{tabular}{|c|c|c|}
    \hline
$\chi$ &  $n_s$    &   $\Delta\psi$  \\
    \hline
$58.05^\circ$ &  2.2805 &   $  0.7312^\circ $\\
$58.06^\circ$ &  2.2804 &   $ 0.7805^\circ$\\
$58.07^\circ$ &  2.2804 &   $0.8179^\circ $\\
$58.08^\circ$ & 2.2803  &    $ 0.9761^\circ $\\
$58.09^\circ$ &  2.2802 &   $ 1.0302^\circ $\\
$58.099^\circ$ &  2.2801  & $1.4^\circ$ \\
\hline
\end{tabular}
\caption{Maximum  $\Delta\psi$ for some choice of parameters.}
\end{center}
\end{table}

\newpage

\pagebreak

\end{document}